\documentclass{jfm}
\usepackage{graphicx}
\usepackage{natbib}
\usepackage{bm,amsmath}

\ifCUPmtlplainloaded \else
  \checkfont{eurm10}
  \iffontfound
    \IfFileExists{upmath.sty}
      {\typeout{^^JFound AMS Euler Roman fonts on the system,
                   using the 'upmath' package.^^J}%
       \usepackage{upmath}}
      {\typeout{^^JFound AMS Euler Roman fonts on the system, but you
                   dont seem to have the}%
       \typeout{'upmath' package installed. JFM.cls can take advantage
                 of these fonts,^^Jif you use 'upmath' package.^^J}%
       \providecommand\upi{\pi}%
      }
  \else
    \providecommand\upi{\pi}%
  \fi
\fi

\ifCUPmtlplainloaded \else
  \checkfont{msam10}
  \iffontfound
    \IfFileExists{amssymb.sty}
      {\typeout{^^JFound AMS Symbol fonts on the system, using the
                'amssymb' package.^^J}%
       \usepackage{amssymb}%
         
         \let\geq=\geqslant
      }{}
  \fi
\fi

\ifCUPmtlplainloaded \else
  \IfFileExists{amsbsy.sty}
    {\typeout{^^JFound the 'amsbsy' package on the system, using it.^^J}%
     \usepackage{amsbsy}}
    {\providecommand\boldsymbol[1]{\mbox{\boldmath $##1$}}}
\fi

\newcommand\Real{\mbox{Re}} % cf plain TeX's \Re and Reynolds number
\newsavebox{\astrutbox}
\sbox{\astrutbox}{\rule[-5pt]{0pt}{20pt}}

\title[Deformation of a polymer in a random flow with long correlation time]%
{Deformation of a flexible polymer in a random flow
with long correlation time}

\author[S. Musacchio and D. Vincenzi]%
{S\ls T\ls E\ls F\ls A\ls N\ls O\ns
M\ls U\ls S\ls A\ls C\ls C\ls H\ls I\ls O
\and D\ls A\ls R\ls I\ls O\ns
V\ls I\ls N\ls C\ls E\ls N\ls Z\ls I}

\affiliation{
CNRS UMR 6621, Laboratoire J.A. Dieudonn\'e, Universit\'e de 
Nice Sophia Antipolis, \\  Parc Valrose, 06108 Nice, France}

\pubyear{??}
\volume{??}
\pagerange{??--??}
\date{?? and in revised form ??}

\newcommand{\Rmax}{R_{\text{max}}}

\begin{document}

\maketitle

\begin{abstract}
The effects induced by long temporal correlations 
of the velocity gradients on the dynamics of a flexible polymer
are investigated by means of theoretical and numerical analysis of 
the Hookean and FENE dumbbell models in a random renewing flow. 
For Hookean dumbbells,
we show that long temporal correlations 
strongly suppress the Weissenberg-number dependence
of the power-law tail 
characterising the probability density function (PDF) 
of the elongation.
For the FENE model, the PDF
becomes bimodal,
and the coil--stretch transition occurs
through the simultaneous drop and rise of the two peaks associated with
the coiled and stretched configurations, respectively. 
\end{abstract}

%\begin{keywords}
%\end{keywords}

\section{Introduction}

The dynamics of a flexible polymer in a moving fluid
depends strongly on the properties of the velocity gradients.
A remarkable example is given by the coil--stretch transition in a
planar extensional flow~$\bm v=(\epsilon x, -\epsilon y)$ \citep*{dG74,PSC97}. 
This flow is characterised
by a direction of pure compression and a direction of pure stretching with
constant strain rate~$\epsilon$. 
%The conformation of the polymer is the result
%of the competition between the drag force exerted by the flow and
%the entropic force restoring the polymer into its
%equilibrium configuration.
%In an extensional flow,
The elongation of the polymer is controlled by the Weissenberg 
number~$\textit{Wi}=\epsilon\tau_p$, where~$\tau_p$ is the longest relaxation
time of the polymer. The value~$\mbox{\textit{Wi}}=1/2$ marks the coil--stretch 
transition:
for~$\mbox{\textit{Wi}}<1/2$ the polymer remains coiled under the action 
of the entropic elastic force,
whereas for~$\mbox{\textit{Wi}}>1/2$ 
the drag force exerted by the flow overcomes the entropic force and
the polymer unravels almost completely.
As a consequence, the probability density of the extension, $p(R)$, 
has a single pronounced peak, whose position depends on~$\mbox{\textit{Wi}}$.
For~$\mbox{\textit{Wi}}<1/2$, the peak is 
in the neighbourhood of the equilibrium size of the polymer, $R_0$;
for~$\mbox{\textit{Wi}}>1/2$,
the peak approaches the maximum length~$\Rmax$.

The ability of a (non-uniform) laminar flow to deform an isolated 
polymer has now been demonstrated for various
flow configurations (see, e.g., 
the reviews by \citealp{L05} and \citealp{S05}).
The corresponding problem
for random flows was first studied
by \cite{L72,L73}, who observed that the strain tensor of an
incompressible random flow always has a direction of stretching,
although such direction fluctuates in time and in space.
Lumley further remarked that the vorticity penalises polymer
stretching, for it prevents the polymer from remaining aligned with the
principal axes of the strain tensor.
He then concluded that random flows can stretch polymers
if the product of the Lagrangian correlation time 
of the strain tensor and the modulus of the maximum eigenvalue
of the deformation tensor
exceeds a critical value proportional to~$\tau_p^{-1}$.
\citet{GS01} confirmed these conclusions by
showing evidences of a 
significant amount of polymer stretching
in a low-Reynolds-number random flow generated by viscoelastic instabilities.

In random flows, the time scale describing the stretching of
line elements is
the reciprocal of the maximum Lyapunov exponent~$\lambda$. Hence
the appropriate definition of the 
Weissenberg number is~$\mbox{\textit{Wi}}=\lambda\tau_p$.
\citet*{BFL00} and \citet{C00} related the deformation of a polymer
in a random flow to the statistics of the stretching rate or,
more precisely, to the entropy function associated with its
probability density function \citep*[e.g.,][]{CPV93}. 
Based on this analysis,
\citet{BFL00} predicted the existence of the coil--stretch transition
for any random flow with positive~$\lambda$, 
and proposed the following explanation.
For an infinitely extensible polymer ($\Rmax\to\infty$),
the stationary probability density function (PDF)
of the extension has a power-law tail: $p(R)\propto R^{-1-\alpha}$
for~$R\gg R_0$, 
where~$\alpha$ depends
on the form of the entropy function.
The exponent~$\alpha$ is positive for~$\mbox{\textit{Wi}}<1/2$, it decreases 
with increasing~$\mbox{\textit{Wi}}$, 
and becomes negative as~$\mbox{\textit{Wi}}$ exceeds~$1/2$.
Consequently, $p(R)$ is normalizable only for~$\mbox{\textit{Wi}}<1/2$, whereas
for~$\mbox{\textit{Wi}}\geq 1/2$ there is unbounded growth of the extension
and the assumption of infinite extensibility becomes inadequate.
This abrupt change in the statistics of the extension 
is interpreted as indicating the coil--stretch transition in 
random flows. The threshold~$\mbox{\textit{Wi}}=1/2$ reproduces
(to within a numerical factor)
Lumley's criterion%
\footnote{\citet{BFL00} defined the Weissenberg number
as~$\mbox{\textit{Wi}}=2\lambda\tau_p$ so that the coil--stretch transition
occurs at~$\mbox{\textit{Wi}}=1$.}.

If the random flow is isotropic and has zero correlation time,
$p(R)$ can be written
explicitly and the behaviour predicted by \citet{BFL00} can be obtained
by direct computation~\citep*{C00,T03,CMV05,MAV05}.
In this particular case, $\alpha=d(\mbox{\textit{Wi}}^{-1}-2)/2$, 
where~$d$ is the spatial dimension of the flow.

A polymer having finite maximum extension ($\Rmax<\infty$) 
reaches a stationary configuration 
at any~$\mbox{\textit{Wi}}$ since the entropic force forbids extensions 
greater than~$\Rmax$. 
According to the analysis of the infinitely extensible case,
$p(R)$ is now expected to
display a power-law behaviour~$R^{-1-\alpha}$
for~$R_0\ll R\ll\Rmax$.
In a short-correlated flow,
the coil--stretch transition results from
the combination of this
power law  and the cutoff at~$R=\Rmax$:
as~$\mbox{\textit{Wi}}$ increases, the intermediate power law raises and
the peak of~$p(R)$ moves from~$R_0$ to extensions
near to~$\Rmax$  \citep{MAV05}.
An important qualitative difference between extensional and random flows 
must nevertheless be emphasised.
In the former case,
$p(R)$ has a pronounced peak at an extension near either~$R_0$ or~$\Rmax$
depending on whether~$\mbox{\textit{Wi}}$ is less or greater than~$1/2$. 
In the latter case,
$p(R)$ has broad tails signaling %, at any~$\mbox{\textit{Wi}}$, 
the coexistence of coiled and stretched configurations
with predominance of either configuration according to 
the value of~$\mbox{\textit{Wi}}$.
In random flows,
the coil--stretch transition is not as sharp as in the extensional case 
because of the random nature of the velocity gradient and the presence of 
vorticity.

The coil--stretch transition and the relation between
the exponent~$\alpha$ and the maximum Lyapunov exponent~$\lambda$
were investigated experimentally by~\citet*{GCS05} and~\citet*{LS10}. 
In these experiments,
the statistics of the extension was measured by
following a fluorescently labelled polymer
in a random velocity field generated by elastic turbulence.
The PDF of polymer extension was also
investigated by means of
direct numerical simulations of the continuum equation for
the polymer conformation tensor
%(Collins-attivo)
\citep*{EKS02,BCM03}
and by means of Brownian Dynamics simulations 
for elastic dumbbells \citep{CMV05,DS06,BMPB10}
or multi-bead chains \citep{WG10}. The results of the simulations support the
picture of polymer dynamics presented above.

The Weissenberg number suffices to determine
the elongation of a polymer  in laminar flows.
By contrast,  
in random flows an additional dimensionless number may influence the
deformation of the polymer, 
namely, the Kubo number~$\mbox{\textit{Ku}}=\lambda\tau_c$, where~$\tau_c$ is  
the Lagrangian correlation time of the velocity gradient.
The case~$\mbox{\textit{Ku}}=0$ has been briefly reviewed above.
In this paper, we examine the effect of a nonzero~$\mbox{\textit{Ku}}$ 
and show that the dynamics at~$\mbox{\textit{Ku}}>1$
differs significantly from the one predicted 
for short-correlated flows. To emphasise the basic physical mechanisms, we
consider a simplified situation. 
The polymer molecule is modelled as an elastic
dumbbell \citep{BHAC77}. This approximation is appropriate
when attention is restricted to the extension
of the polymer~\citep{WG10}.
As for the random advection, we consider
a two-dimensional linear renewing flow. In renewing flows, time is split
into intervals of length~$\tau_c$ and the velocities in different intervals
are independent and identically distributed \citep[e.g.,][p. 320]{CG95}. 
Linear renewing flows were used by \citet{ZRMS84} to study
the kinematic dynamo effect.
The magnetic
field is actually stretched by the velocity gradient
in the same way as an infinitely 
extensible polymer. The use of a linear renewing flow enables us to
obtain semi-analytical results and to accurately compute the statistics
of polymer extension with moderate numerical effort.

The remainder of the paper is divided as follows.
In section~\ref{sec:dumbbell}, we briefly review the elastic dumbbell model.
In section~\ref{sec:renewing}, we introduce the renewing random flow.
The results are presented in section~\ref{sec:results}. 
Finally, some conclusions are drawn in section~\ref{sec:concl}.

\section{Elastic dumbbell model}
\label{sec:dumbbell}

An elastic dumbbell is composed of two beads joined by a spring.
The beads represent the two ends of the polymer; the spring models
the entropic force.  For the sake of simplicity, we assume that the
flow transporting the dumbbell is two-dimensional.

The vector separating the ends of the polymer,
$\bm R$, satisfies the stochastic ordinary
differential equation \citep[e.g.,][]{BHAC77}:
\begin{equation}
\label{eq:dumbbell}
\dot{\boldsymbol{R}}=\sigma(t)\boldsymbol{R}
-\frac{1}{2\tau_p}\,f(R)\boldsymbol{R}
+\sqrt{\frac{R_0^2}{\tau_p}}\,\bm\xi(t),
\end{equation}
where~$R=\vert\bm R\vert$ 
and~$\bm\xi(t)$ is white noise, i.e., a Gaussian process with
zero mean and two-time
correlation $\langle\xi_i(t)\xi_j(t')\rangle=\delta_{ij}
\delta(t-t')$.
The three terms on the right-hand-side of
equation~\eqref{eq:dumbbell} result from the drag force, the entropic elastic force,
and thermal noise, respectively. 
The $2\times 2$ matrix~$\sigma(t)$ is the velocity gradient evaluated
along the trajectory of the centre of mass of the dumbbell:
$\sigma_{ij}(t)=\partial_j v_i(t)$.
The function~$f(R)$ is identically equal to~$1$ for
an infinitely extensible polymer (Hookean model)
and has the form~$f(R)=1/(1-R^2/\Rmax^2)$ for a finitely extensible 
polymer with nonlinear elasticity
(FENE model). In the former case, $\bm R$ is defined on~$\mathbb{R}^2$;
in the latter case, $\bm R$ belongs to~$[0,\Rmax)\times[0,\Rmax)$.

Equation~\eqref{eq:dumbbell} holds under some assumptions on the
dynamics of the beads.
The velocity field is assumed to be linear at the size of the dumbbell.
The drag on a bead is given by the Stokes law.
Furthermore, hydrodynamic interactions between the beads 
and inertial effects are disregarded.

\section{Linear renewing flow}
\label{sec:renewing}

According to the assumptions of the dumbbell model,
in the reference frame of the centre of mass of the polymer
the velocity field is of the form: $v_i(\bm r,t)=\sigma_{ij}(t)r_j$.
The velocity gradient~$\sigma(t)$ is constant
in each of the time
intervals~$I_n=[n\tau_c,(n+1)\tau_c)$, $n\in\mathbb{N}$.
We denote by~$\sigma_n$ the (random) value taken by~$\sigma(t)$ in~$I_n$:
$\sigma(t)=\sigma_n$ for all~$t\in I_n$.
The random matrices~$\sigma_n$ are identically
distributed and statistically independent. We assume
that, for a fixed ~$n$, 
$\sigma_n$ is Gaussian, zero-mean,
statistically isotropic, and traceless
(so that the velocity field is incompressible). As a result, $\sigma(t)$
takes the form:
\begin{equation}
\label{eq:renew}
\sigma(t)=\frac{S}{2}
\begin{pmatrix}
\zeta_1(t) & \zeta_2(t)
\\
\zeta_2(t) & -\zeta_1(t)
\end{pmatrix}
+\frac{\varOmega}{\sqrt{2}}
\begin{pmatrix}
0 & \zeta_3(t)
\\
-\zeta_3(t) & 0
\end{pmatrix}
\end{equation}
with~$S$ and~$\varOmega$
positive constants. The elements~$\zeta_i(t)$ 
satisfy: $\zeta_i(t)=\zeta_{i,n}$
for all~$t\in I_n$, where the~$\zeta_{i,n}$ are Gaussian random variables
such that
$\langle\zeta_{i,n}\rangle =0$ for all~$i$ and~$n$ and
$
\langle\zeta_{i,n}\zeta_{j,m}\rangle=\delta_{ij}\delta_{nm}
$
for all~$i,j=1,2,3$ and~$n,m\in\mathbb{N}$.
The mean-squared strain and rotation rates are~$S^2$ and~$\varOmega^2$, respectively.
We set~$\varOmega=S$ to reproduce the relation 
holding for the solution of the Navier--Stokes
equation \citep[e.g.,][p.~20]{F95}.

Strictly speaking, the flow considered is not
statistically stationary in time. Nevertheless, over time ranges longer
than~$\tau_c$, it can be considered as a good 
approximation to a stationary flow with correlation time~$\tau_c$
owing to the invariance with respect to the transformation $t\to t+\tau_c$ and 
thanks
to the independence of the matrices~$\sigma(t)$  in different time 
intervals~$I_n$
\citep{ZRMS84}.
The $\delta$-correlated flow (i.e. white-in-time noise) 
is recovered by letting~$\tau_c$ tend to zero while holding~$S^2\tau_c$ constant.

As mentioned in the introduction, 
the elongation of a polymer is
related to the statistics of the stretching rate of the flow.
For a review on the
entropy function and the generalized Lyapunov 
exponents in statistical physics, we
refer the reader to the book by \cite{CPV93}. Here, we briefly
recall some basic elements of the theory.
If $\bm\ell(t)$ denotes a fluid-line element, the stretching rate at
time~$t$ is defined as
$
\gamma(t)=t^{-1}\ln[\ell(t)/\ell(0)].
$
The maximum Lyapunov exponent is the long-time limit of the average
stretching rate: $\lambda=\lim_{t\to\infty}\langle\gamma(t)\rangle$,
where the average is taken over the realizations of~$\sigma(t)$.
The PDF of~$\gamma$ for long~$t$ takes the 
large-deviation form \citep{CPV93}:
\[
P(\gamma,t)\propto \mathrm{e}^{-G(\gamma)t},
\]
where $G(\gamma)$ is the entropy function. $G(\gamma)$
is non-negative
and attains its minimum value at the point~$\lambda$.
Equivalently, the stretching properties of the flow can be 
characterised by
the generalized Lyapunov exponents,
defined as the
rate of growth of the moments of~$\ell(t)$:
\[
L(q)=\lim_{t\to\infty}\frac{1}{t}\ln\left\langle\left( 
\dfrac{\ell(t)}{\ell(0)}\right)^q\right\rangle.
\]
The function $L(q)$ is the Legendre transform of~$G(\gamma)$,
$L(q)=\max_\gamma[\gamma q-G(\gamma)]$, and
satisfies~$L'(0)=\lambda$.
For small~$q$, $L(q)$ can be obtained by using the quadratic approximation
$G(\gamma)\approx(\gamma-\lambda)^2/(2\Delta)$ so that
$L(q)\approx\lambda q+\Delta q^2/2$. 

\begin{figure}
\setlength{\unitlength}{1cm}
\centering
\includegraphics[width=10cm]{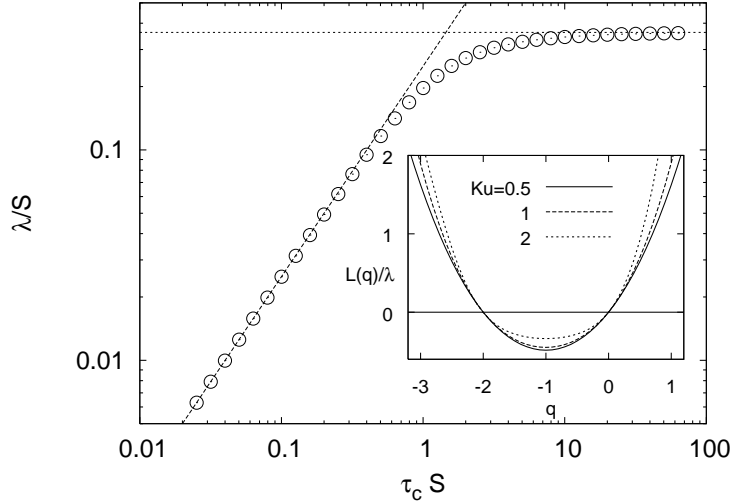}
\caption{Lyapunov exponent as a function of~$\tau_c$ (circles). 
Dashed and dotted lines represent the asymptotic behaviours~\eqref{eq:lyap_asymptotics}
for $\tau_c\to 0$ and $\tau_c\to \infty$, respectively.
Inset: Generalized Lyapunov exponents $L(q)$ for three values of $Ku$. 
Note that $L(-d)=L(0)=0$ in agreement with the properties of the generalized Lyapunov
exponents proved by \cite{ZRMS84}.}
\label{fig:lyap}
\end{figure}

For the linear renewing flow, we can compute~$L(q)$ by adapting
to the present problem 
the method described by \cite{GB92}
\cite[see also][pp.~322--326]{CG95}.
A line element satisfies the equation:
\begin{equation}
\label{eq:element}
\dot{\bm\ell}(t)=\sigma(t)\bm\ell(t).
\end{equation}
Since $\sigma(t)$ is constant in~$I_n$,
the solution of
equation~\eqref{eq:element} for~$t\in I_n$ is:
$
\bm\ell(t)=\mathrm{e}^{\sigma_n t}\bm\ell(n\tau_c).
$
Hence
\[
\bm\ell((n+1)\tau_c)=\mathsfbi{J}_n \bm\ell(n\tau_c),
\]
where $\mathsfbi{J}_n=\mathrm{e}^{\sigma_n \tau_c}$ is written
\[
\mathsfbi{J}_n=\dfrac{1}{2\omega_n}
\begin{pmatrix}
2\omega_n c_n^++S\zeta_{1,n}c_n^-
&
(S\zeta_{2,n}+\sqrt{2}\varOmega\zeta_{3,n})c_n^-
\\[5mm]
(S\zeta_{2,n}-\sqrt{2}\varOmega\zeta_{3,n})c_n^-
&
2\omega_n c_n^+ -S\zeta_{1,n}c_n^-
\end{pmatrix}
\]
with $\omega_n=\frac{1}{2}\sqrt{S^2(\zeta_{1,n}^2+\zeta_{2,n}^2)-2\varOmega^2\zeta_{3,n}^2}$, 
$c_n^+=(\mathrm{e}^{\omega_n\tau_c}+\mathrm{e}^{-\omega_n\tau_c})/2$, 
and~$c_n^-=(\mathrm{e}^{\omega_n\tau_c}-\mathrm{e}^{-\omega_n\tau_c})/2$
(we remind the reader that $\zeta_{i,n}$ is the constant
value taken by the random function $\zeta_i(t)$ 
in~$I_n$).
If $\omega_n$ is real, then~$\mathsfbi{J}_n$ can be written in terms of
hyperbolic functions; for a non-real complex~$\omega_n$, $\mathsfbi{J}_n$
involves trigonometric functions.
By using the statistical isotropy of the flow and the fact that the 
matrices~$\sigma_n$ are identically distributed and independent
for different~$n$, it is possible to 
show the following relation \citep{GB92,CG95}:
\begin{equation}
\label{eq:iteration}
\frac{\langle\ell^q((n+1)\tau_c)\rangle}{\langle\ell^q(n\tau_c)\rangle}=
\left\langle\vert \mathsfbi{J}_n\bm e\vert^{q}\right\rangle_{I_n},
\end{equation}
where~$\bm e$ is any constant unit vector 
and the average $\langle\cdot\rangle_{I_n}$ is taken over 
the statistics of the gradient in the interval~$I_n$ only.
Iterating equation~\eqref{eq:iteration} yields: $\langle \ell^q(n\tau)\rangle=
\langle \rho^q\rangle^n_{I_n} \ell^q(0)$
with~$\rho^2\equiv \vert \mathsfbi{J}_n
\bm e\vert^2=\bm e^\mathrm{T}\mathsfbi{J}^\mathrm{T}_n\mathsfbi{J}_n\bm e$. 
It follows that~$L(q)$ and~$\lambda$ can be written as
\begin{equation}
\label{eq:generalized}
L(q)=\tau_c^{-1}\,\ln\langle \rho^q\rangle_{I_n} %\qquad q>0,
\qquad \text{and}\qquad
\lambda=\tau_c^{-1}\,\langle\ln \rho\rangle_{I_n}.
\end{equation}
Given that the~$\sigma_n$ are identically distributed, the average
can be taken over any interval~$I_n$.
Equations~\eqref{eq:generalized} provide a simple way to compute 
the function~$L(q)$ and the Lyapunov exponent.
The behaviour of~$\lambda$ as a function of~$\tau_c$ is reported
in figure~\ref{fig:lyap}.
%%%%%%%%%%%%%%%%%%%%%%%%%%%%%%%%%%%%%%%%%%%
%%%%%%%%%%%%%%%%%%%%%%%%%%%%%%%%%%%%%%%%%%
The following asymptotic behaviours hold \citep{CFKL96}:
%\footnote{To reproduce the same notation as in the paper
%by \cite{CFKL96}, $S$ and~$\varOmega$ must be multiplied by~$2$
%and by~$\sqrt{2}$, respectively.}:
\begin{equation}
\label{eq:lyap_asymptotics}
\lambda \sim \frac{S^2 \tau_c}{4} \qquad (\tau_c\to 0)
\qquad
\text{and} 
\qquad
\lambda\sim \Real\langle\omega_n\rangle_{I_n}=
\sqrt{\frac{\upi}{2}}\, \frac{S}{2} \sqrt{\frac{S^2}{S^2+2\varOmega^2}}
\qquad (\tau_c\to \infty).
\end{equation}
For $\varOmega=S$, one obtains $\lambda\sim\sqrt{\upi/6}\,S/2$
as $\tau_c\to \infty$.
Thus the Lyapunov exponent becomes independent of~$\tau_c$
for large~$\tau_c$, and the convergence to the asymptotic value
is exponential.
By contrast, the variance~$\Delta$ monotonically increases
like~$\Delta\sim S^2\tau_c/4$ both for small and large~$\tau_c$ \citep{CFKL96}, 
and therefore the
generalized Lyapunov exponents do not saturate
to a constant value (figure~\ref{fig:lyap}).
We stress the fact that the above asymptotic behaviours 
hold for any two-dimensional random flow that is incompressible
and statistically isotropic and not only for the renewing flow.

%%%%%%%%%%%%%%%%%%%%%%%%%%%%%%%%%%%%%
\section{Statistics of polymer extension}
\label{sec:results}
In order to investigate the influence of temporal correlations 
of the velocity gradients on the dynamics of polymers, we
numerically integrated equation~\eqref{eq:dumbbell} for 
the elastic dumbbell model, where the velocity gradients 
were determined by equation~\eqref{eq:renew}. 
We computed the PDF
of the polymer elongation~$R$ for various values 
of~$\mbox{\textit{Ku}}$ and~$\mathit{Wi}$.
The numerical integration has been performed using the 
predictor--corrector scheme proposed
by  \cite{O96} for the FENE model and 
the stochastic Runge--Kutta algorithm introduced by \cite{H92}
for the Hookean model.
The statistics of polymer elongation has been computed by following 
the dynamics of a single dumbbell for $10^7$ time intervals~$I_n$. 

%%%%%%%%%%%%%%%%%%%%%%%%%%%%%%%%%%%%%%%%%%%
\begin{figure}
\setlength{\unitlength}{1cm}
\centering
\includegraphics[width=10cm]{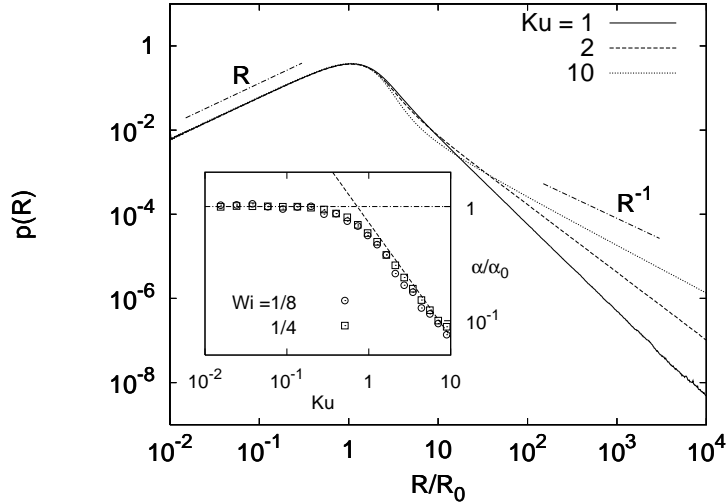}
\caption{PDFs of polymer elongation, $p(R)$, 
for the Hookean model at $\mathit{Wi}=1/4$. 
Note the power-law behaviours $p(R)\propto R^{d-1}$ for $R \ll R_0$ 
and $p(R)\propto R^{-1-\alpha}$ for $R \gg R_0$. 
Inset: Exponent $\alpha$ as a function of $\mathit{Ku}$. 
Here $\alpha_0 = \lim_{\mathit{Ku}\to 0}\alpha$. 
Dash-dotted and dashed lines are 
the asymptotic behaviours~\eqref{eq:alpha} and~\eqref{eq:asympt} 
for $\mathit{Ku} \to 0$ and $\mathit{Ku} \to \infty$, respectively.}
\label{fig:linear}
\end{figure}
%%%%%%%%%%%%%%%%%%%%%%%%%%%%%%%%%%%%%%%%%%

Within the Hookean model, the PDF of the 
extension behaves 
like~$p(R)\propto R^{-1-\alpha}$
for $R\gg R_0$ \citep{BFL00}.
The coil--stretch transition is identified by 
the change of sign of $\alpha$, which occurs at $\mathit{Wi} = 1/2$.
The effects of the temporal correlation of the velocity gradients 
can be quantified through the dependence of~$\alpha$ on the Kubo number. 
The numerical simulations of the Hookean model show 
that for small values of~$\mathit{Ku}$ 
the tail of~$p(R)$ is almost independent of~$\mathit{Ku}$; 
conversely, for $\mathit{Ku} \gtrsim 1$, the tail
raises as~$\mathit{Ku}$ increases and approaches the slope~$-1$
(figure~\ref{fig:linear}).
Because of long temporal correlations, even polymers with a short relaxation time 
occasionally experience significant stretching events, and 
this effect produces a power-law tail 
close to the coil--stretch transition also for small values of~$\mathit{Wi}$.  

This intuitive idea can be rationalized by the following argument.  
The exponent~$\alpha$ determines the lowest order such 
that~$\langle R^\alpha\rangle$ diverges
and satisfies~$\alpha=2\tau_p L(\alpha)$ \citep{BCM03}.
For $\mathit{Wi}$ near to 1/2, $\alpha$ is not far from zero and it is 
appropriate to use the quadratic approximation for~$L(q)$.
The equation~$\alpha=2\tau_p L(\alpha)$ then yields \citep{BFL00}:
\begin{equation}
\label{eq:alpha}
\alpha=\frac{\lambda}{\Delta}\left(\frac{1}{\mathit{Wi}}-2\right).
\end{equation}
While the dependence of~$\alpha$ on the Weissenberg number is contained entirely
in the second factor, the Kubo number enters the expression for~$\alpha$ through the
ratio~$\lambda/\Delta$. For the linear renewing flow, 
the asymptotic behaviours discussed in section~\ref{sec:renewing} give:
\begin{equation}
\label{eq:asympt}
\lambda/\Delta\sim 1 \quad (\mathit{Ku}\to 0)
\qquad
\text{and}
\qquad
\lambda/\Delta=O(\mathit{Ku}^{-1}) \quad (\mathit{Ku}\to\infty).
\end{equation}
One obtains that for small values of~$\mathit{Ku}$ the tail of~$p(R)$ 
is independent of~$\mathit{Ku}$, whereas
$\alpha=O(\mathit{Ku}^{-1})$ as $\mathit{Ku}$ is increased, 
in agreement with our numerical findings. 
%%%%%%%%%%%%%%%%%%%%%%%%%%%%%%%%%%%%%%%%%%%
\begin{figure}
\setlength{\unitlength}{1cm}
\centering
\includegraphics[width=6.7cm]{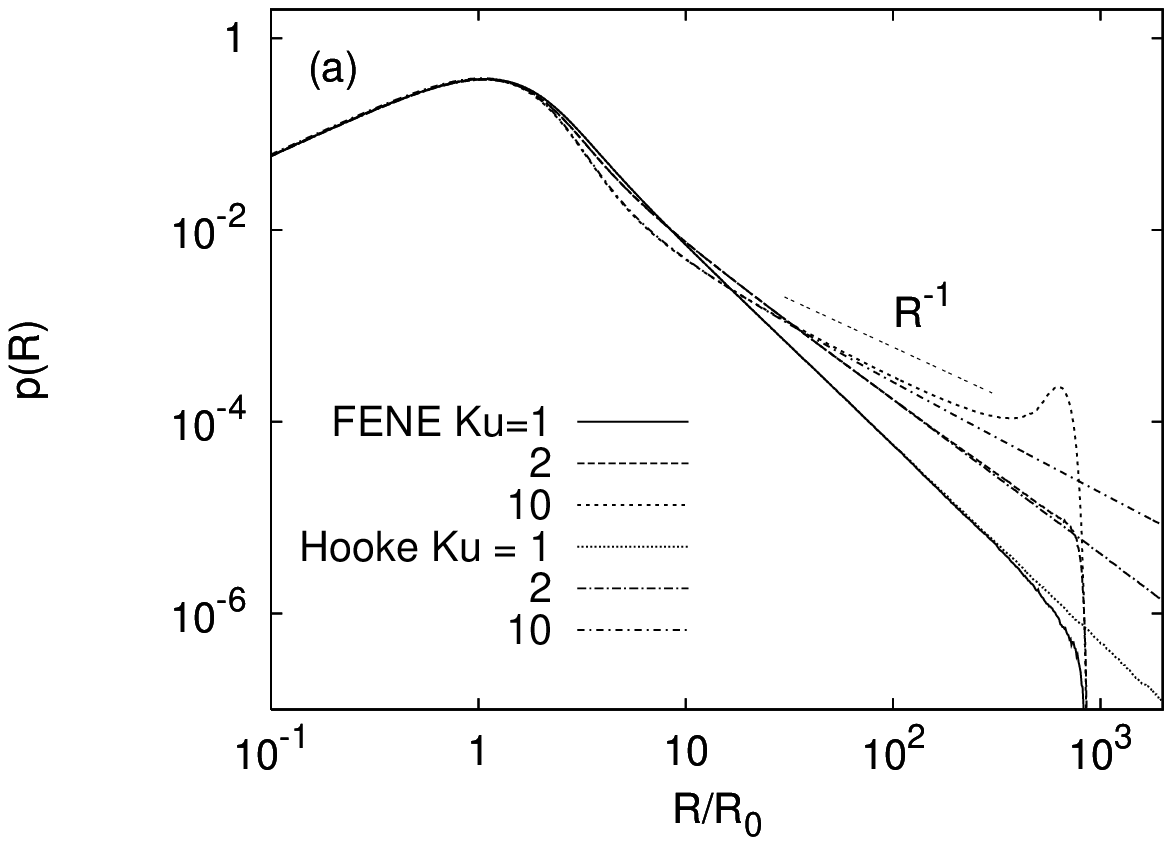}
\hfill
\includegraphics[width=6.7cm]{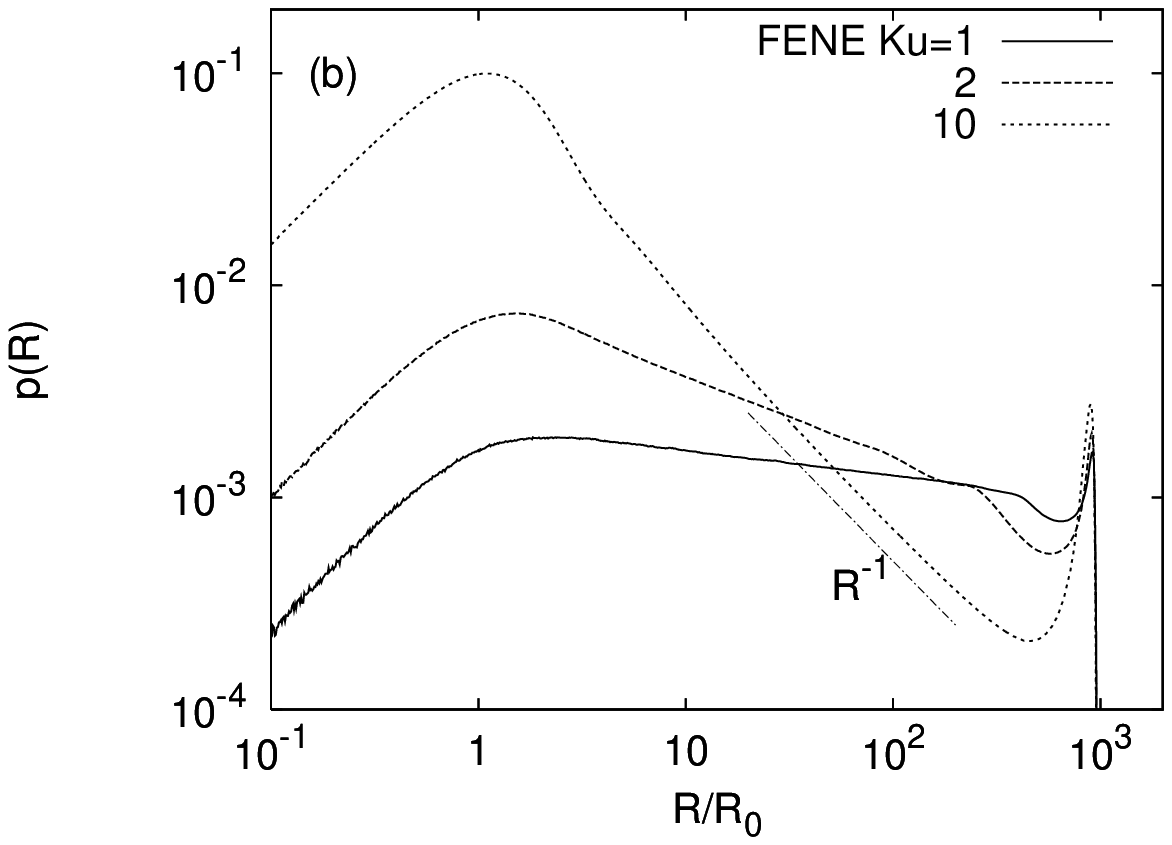}
\caption{
PDFs of polymer elongation, $p(R)$ for the FENE model
at various values of~$\mathit{Ku}$ 
for $\mathit{Wi}=1/4$ (panel $(a)$) 
and $\mathit{Wi}=1$ (panel $(b)$).
The maximum extension is set to~$\Rmax= 10^3 R_0$.}
\label{fig:r1000}
\end{figure}
%%%%%%%%%%%%%%%%%%%%%%%%%%%%%%%%%%%%%%%%%%

\begin{figure}
\setlength{\unitlength}{1cm}
\centering
\includegraphics[width=8.5cm]{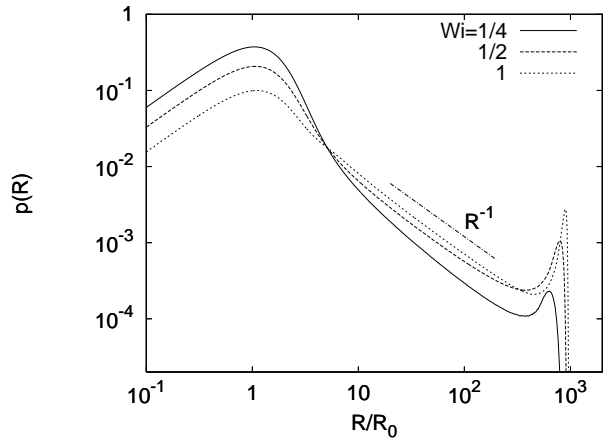}
\caption{
PDFs of polymer elongation for the FENE model 
at various values of $\mathit{Wi}$ 
in a long-correlated flow ($\mathit{Ku}=10$).  
The maximum extension is set to~$\Rmax= 10^3 R_0$.}
\label{fig:r1000_ku10}
\end{figure}

Although for an infinitely extensible polymer $p(R)$ is normalizable 
only for~$\alpha>0$ (equivalently for~$\mathit{Wi}>1/2$), the calculation leading
to the power-law behaviour remains formally valid also if~$\alpha<0$ ($\mathit{Wi}>1/2$).
In this case, the stationary PDF of the extension
exists only if
the nonlinearity of the elastic force is taken into account, 
and the power-law prediction is expected to hold for intermediate
extensions~$R_0\ll R\ll \Rmax$. 
To allow the 
development of the intermediate power law,
we performed numerical simulations of the FENE model 
at artificially high~$\Rmax/R_0$. 
%To address this issue we performed numerical simulations of the FENE model, 
%at artificially high~$\Rmax/R_0$, in order to allow the 
%development of the power-law tail in the range $R_0\ll R\ll \Rmax$.
For $\mathit{Wi} < 1/2$ the results agree with those 
obtained for the Hookean model (figure~\ref{fig:r1000}, panel~(a)). 
For  $\mathit{Wi} > 1/2$ an increase of~$\mathit{Ku}$ produces a decrease
of the intermediate slope of~$p(R)$, which can be negative even for very 
large~$\mathit{Wi}$ (figure~\ref{fig:r1000}, panel~(b)).
According to equation~\eqref{eq:asympt}, near~$\mathit{Wi}=1/2$
the range of variation of~$\alpha$ as a function 
of~$\mathit{Wi}$ can be made arbitrarily small by increasing~$\mathit{Ku}$, 
to the extent that the intermediate slope of~$p(R)$ becomes almost
independent of~$\mathit{Wi}$ (figure~\ref{fig:r1000_ku10}).
%%%%%%%%%%%%%%%%%%%%%%%%%%%%%%%%%%%%%%%%%%%
%%%%%%%%%%%%%%%%%%%%%%%%%%%%%%%%%%%%%%%%%%
Our findings show that the picture 
of a coil--stretch transition associated with
the increase of the intermediate slope of~$p(R)$
does not hold in a long-correlated flow. At large~$\mathit{Ku}$, the appearance
of the stretched state occurs through 
the drop of the maximum at~$R\approx R_0$
and the simultaneous rise of a second maximum 
near~$\Rmax$ (figure~\ref{fig:r1000_ku10}).

\begin{figure}
\setlength{\unitlength}{1cm}
\centering
\includegraphics[width=8.5cm]{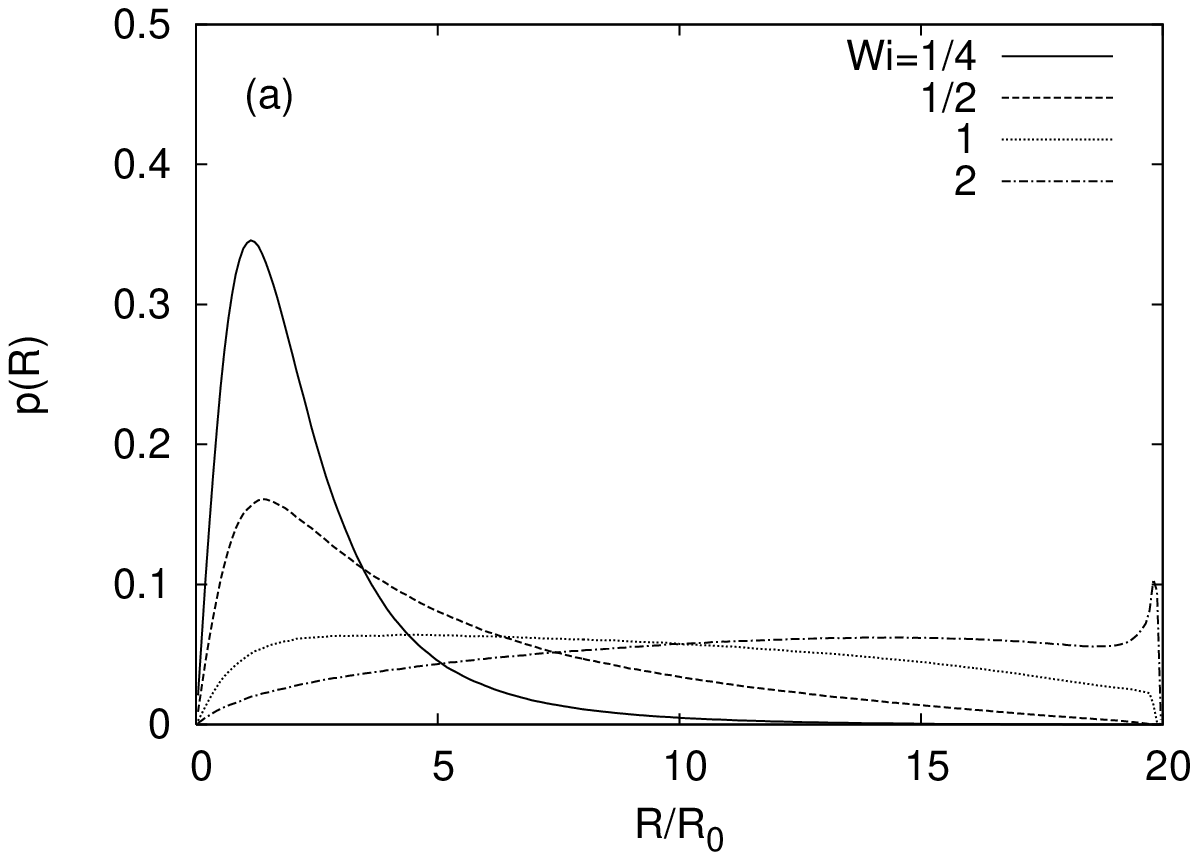}
\includegraphics[width=8.5cm]{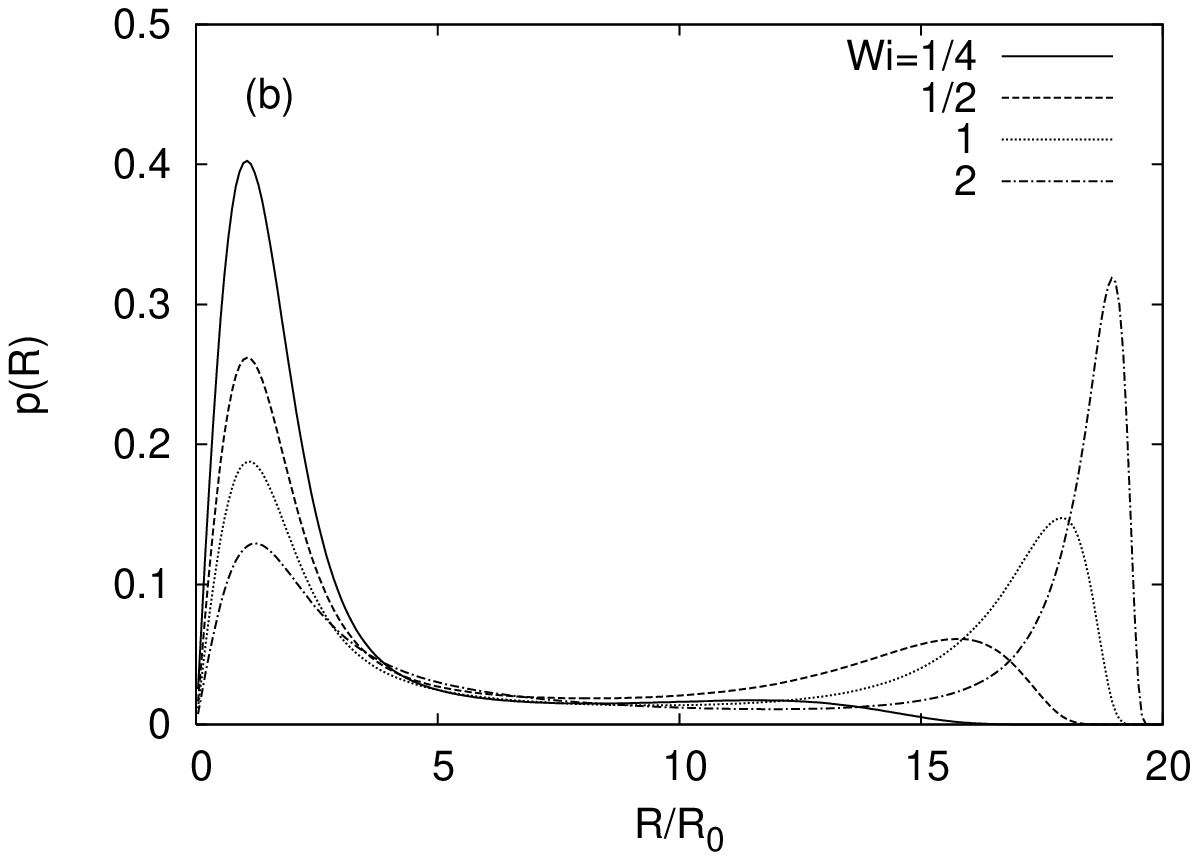}
\caption{
PDFs of polymer elongation, $p(R)$, for the FENE model, 
at various $\mathit{Wi}$ numbers, 
in a short-correlated flow ($\mathit{Ku}=10^{-2}$, panel $(a)$) 
and in a long-correlated flow ($\mathit{Ku}=10$, panel $(b)$). 
The maximum extension is fixed to $\Rmax= 20 R_0$.}
\label{fig:Ku-r20}
\end{figure}

For realistic values of~$\Rmax/R_0$, 
it is difficult to detect a clean power law at intermediate extensions
and the above scenario becomes even more relevant. 
The numerical simulations of the FENE model for $\Rmax/R_0=20$ show that, 
for short-correlated flows ($\mathit{Ku} \ll 1$), 
the coil--stretch transition manifests through the
gradual widening of~$p(R)$ and the displacement of its maximum from values 
close to~$R_0$  towards values close to~$\Rmax$ 
(see figure~\ref{fig:Ku-r20}, panel (a)).
This behaviour is in agreement with the theoretical prediction for the
$\delta$-correlated isotropic flow \citep{MAV05} 
and with recent experimental measurements 
of polymer elongation in a random shear flow \citep{LS10}.  
For long-correlated flows ($\mathit{Ku} \gtrsim 1$), 
the dependence of~$p(R)$ on~$\mathit{Wi}$ 
is very different. In this case,
increasing~$\mathit{Wi}$ at fixed~$\mathit{Ku}$ rather produces 
the drop of the peak at~$R_0$
and the simultaneous formation of a second peak near~$\Rmax$
(see figure~\ref{fig:Ku-r20}, panel (b)).
The PDF of the extension is bimodal and the
two maxima are clearly distinct: intermediate extensions are much less
likely than in the small-$\mathit{Ku}$ case.

The shape of~$p(R)$ for large~$\mathit{Ku}$ can be explained
intuitively as follows. If the Lagrangian correlation time of the velocity
gradient is long, a typical trajectory of a polymer is composed of
long portions where the gradient can be thought as frozen with a fixed
direction of stretching. Some of those portions will be characterised
by a weak stretching rate, some by a strong one. 
The response of the polymer
to the velocity gradient along one of those portions of the trajectory 
will be similar to the response that a polymer would have in an extensional flow 
with strain rate comparable to the stretching rate. As mentioned earlier,  
the PDF of~$R$ in an extensional flow has a sharp peak either
at small extensions or at long extensions depending on the strain rate, and
intermediate extensions are very unlikely. 
Therefore, the evolution of an isolated polymer at large~$\mathit{Ku}$
consists of a sequence of long deformation events, 
whose intensity can produce either a coiled or a highly stretched extension, 
whence the bimodal shape of~$p(R)$. 
%The two maxima are obviously not as sharp as for a real extensional flow 
%because of the random behaviour of the velocity gradient.
%%%%%%%%%%%%%%%%%%%%%%%%%%%%%%%%%%%%%%%%%%%
%%%%%%%%%%%%%%%%%%%%%%%%%%%%%%%%%%%%%%%%%%

%%%%%%%%%%%%%%%%%%%%%%%%%%%%%%%%%%%
\section{Conclusions}
\label{sec:concl}
We have studied the influence of long temporal correlations on 
the dynamics of a Hookean dumbbell
and of a FENE dumbbell in a two dimensional 
random renewing flow.
The time correlation of this simple model flow can be changed 
arbitrarily, and this property
enabled us to obtain semi-analytic predictions 
for the Kubo-number dependence of the statistics of polymer elongations. 
It is known that the PDF of the elongation of Hookean dumbbells is characterised
by a power-law tail and that this power law also describes the behavior
of the PDF of FENE dumbbells at intermediate elongations.
We have shown that
in the long-correlated limit ($\mathit{Ku} \to \infty$) 
the power-law tail 
becomes almost insensitive to~$\mathit{Wi}$ 
and its slope is not an effective indicator for the coil--stretch transition. 
In the case of the FENE model, the long temporal correlations strongly 
affect the shape of the PDF, which shows two distinct maxima corresponding 
to the coiled and stretched configurations. 
For $\mathit{Ku} \gg 1$ we have found a new scenario 
for the coil--stretch transition, which occurs through 
the drop of the peak at~$R_0$
and the simultaneous rise of the second peak near~$\Rmax$.

Our findings can be understood in terms of basic properties of 
long-correlated flows, and the underlying mechanisms does not rely 
on the peculiar choice of the model flow used in our study. 
It is therefore arguable that the phenomena discussed here could be 
observed also in realistic flows. 
In particular, it would be interesting to investigate by means of 
numerical simulations or experimental measurements whether 
the presence of long-lived structures in high-Reynolds-number flows 
could result in the appearence of the 
two-peak coil--stretch transition scenario depicted here. 

\begin{acknowledgements}
We are grateful to Antonio Celani and Prasad Perlekar for 
fruitful discussions.
\end{acknowledgements}

\end{document}